\documentclass[
    ,final            % use final for the camera ready runs
  ]
  {aipproc}

\layoutstyle{6x9}

\def\Journal#1#2#3#4{{#1} {\bf #2}, #3 (#4)}
 
\def\NPA{Nucl.\ Phys.\ A}
\def\NPB{Nucl.\ Phys.\ B}
\def\PLB{Phys.\ Lett.\ B}
\def\PRL{Phys.\ Rev.\ Lett.}
\def\PRD{Phys.\,Rev.\,D}
\def\PRC{Phys.\,Rev.\,C}

\def\xbj{x_{Bj}}

\begin{document}

\title{Neutral Pion Suppression at Forward Rapidities in d+Au 
Collisions at STAR}


\author{G.~Rakness for the STAR Collaboration}{
  address={Penn State University/Brookhaven National Lab}
}

\begin{abstract}
Measurements of the inclusive yields of $\pi^0$ mesons in p+p and 
d+Au collisions at center of mass energy $\sqrt{s_{NN}}=200\,$GeV 
and pseudorapidity $\langle\eta\rangle=4.00$ (d beam direction) 
are reported.
The yield for p+p collisions is in general agreement with 
perturbative QCD calculations.
The d+Au yield is in agreement with a calculation which models 
the Au nucleus as a Color Glass Condensate for forward particle
production.
The nuclear modification factor derived from the inclusive 
yields is qualitatively consistent with models which suppress
the gluon density in nuclei.
\end{abstract}

\maketitle

%%%%%%%%%%%%%%%%%%%%%%%%%%%%%%%%%%%%%%%%%%%%
%\section{Introduction}
%%%%%%%%%%%%%%%%%%%%%%%%%%%%%%%%%%%%%%%%%%%%

In contrast to the nucleon, very little is known about the density
of gluons in nuclei \cite{hirai}.
For protons, the gluon parton distribution function (PDF) is
constrained primarily by scaling violations in deeply-inelastic
lepton scattering (DIS) measured at the HERA collider \cite{hera}.
The data are accurately described by QCD evolution equations that
allow the determination of the gluon PDF.
As the momentum fraction of the parton ($\xbj$) decreases,
the gluon PDF is found to increase.
At sufficiently low $\xbj$, the increase in gluon splitting is 
expected to become balanced by gluon-gluon recombination, 
resulting in gluon saturation.
In nuclei, the density of gluons per unit transverse area is
expected to be larger than in nucleons, and so is a natural 
environment in which to establish if, and under which conditions, 
gluon saturation occurs.
Quantifying if gluon saturation occurs at RHIC energies is 
important because most of the matter created in heavy ion 
collisions is expected to evolve from an initial state produced 
by the collisions of low-$\xbj$ gluon fields in the nuclei 
\cite{gyulassymclerran}.
Fixed target nuclear DIS experiments are restricted in the 
kinematics available and have determined the nuclear gluon PDF 
for $\xbj\stackrel{>}{_\sim}0.02$ \cite{hirai}.
%have reported a suppression of
%the inclusive structure function normalized to proton DIS at
%low-$\xbj$ \cite{fixedtarget}, but 
In a conventional pQCD description of d(p)+A collisions at 
$\sqrt{s_{NN}}=200\,$GeV, inclusive forward hadroproduction 
probes the nuclear gluon 
density over a broad distribution of $\xbj$ peaked around 0.02
and extending well below $\xbj\approx 0.005$ \cite{gsv}.

Using factorization in a perturbative QCD (pQCD) framework, the
PDF's and fragmentation functions (FF's) measured in
electromagnetic processes can be used in the description of
hadronic scattering processes.
In p+p collisions at $\sqrt{s}=200\,$GeV, factorized leading
twist pQCD calculations have been shown to quantitatively describe
inclusive $\pi^0$ production over a broad rapidity
window \cite{STARpi0,PHENIXpi0}.
In pQCD, forward $\pi$ production in p+p collisions probes 
low-$\xbj$ gluons (g) in one proton using the valence quarks (q) of 
the other.
Recently, the yield of forward negatively charged hadrons
($h^-$) in the d-beam direction of d+Au collisions was found to
be reduced when normalized to p+p collisions \cite{BRAHMS}.
The reduction is especially significant since isospin effects
are expected to suppress $h^-$ production in p+p collisions, but not
in d+Au collisions \cite{gsv}.

Many models attempt to describe forward hadroproduction from nuclear 
targets.
Saturation models \cite{saturation} include a QCD based theory
called the Color Glass Condensate (CGC) \cite{cgc,cgc-cross-section}.
Another approach models quarks scattering coherently from multiple
nucleons, leading to an effective shift in the $\xbj$
probed \cite{coherent}.
Shadowing models suppress the nuclear gluon PDF in a standard
factorization framework \cite{shadowing}.
Parton recombination models modify the fragmentation of a quark
passing through many gluons \cite{recombination}.
Other descriptions include factorization breaking in heavy
nuclei \cite{factorization}.
More data are needed to constrain the mechanisms for forward 
hadroproduction in nuclear collisions.

In this paper we present the inclusive yields of high energy 
$\pi^0$ mesons at $\langle\eta\rangle=4.00$ 
($\eta = -\ln[\tan(\theta/2)]$) in p+p and d+Au collisions
at $\sqrt{s_{NN}}=200\,$GeV.
The results are compared with theoretical predictions.
The $\eta$ dependence of the normalized cross section ratio is 
also presented.
Analysis of the azimuthal correlations of the forward $\pi^0$ with
coincident hadrons at midrapidity is presented elsewhere 
\cite{dis2004}.

%%%%%%%%%%%%%%%%%%%%%%%%%%%%%%%%%%%%%%%%%%%%%%%%%%%%%%%%%%%%%%%%%%%
%\section{Experimental Results}
%%%%%%%%%%%%%%%%%%%%%%%%%%%%%%%%%%%%%%%%%%%%%%%%%%%%%%%%%%%%%%%%%%%

%One of its principal components is a time projection chamber used for
%tracking charged particles produced at $|\eta|<1.2$.
A forward $\pi^0$ detector (FPD) 
%comprising $7\times 7$ matrices of
%$3.8\times 3.8\times 45\,$cm$^3$ Pb-glass detectors 
was installed at the Solenoidal Tracker At RHIC (STAR) 
to detect high energy $\pi^0$ mesons with $3.3< \eta < 4.1$.
In the 2002 run, p+p collisions at $\sqrt{s}=200\,$GeV were 
studied with a prototype FPD \cite{STARpi0}.
In the 2003 run, p+p collisions were studied with the complete FPD
and exploratory measurements were performed for d+Au collisions
at $\sqrt{s_{NN}}=200\,$GeV.
Details of the FPD performance can be found in 
Ref.~\cite{moriond05}.
The luminosity was determined by measuring the transverse
size of the colliding beams and the number of colliding ions, 
giving a normalization error of $\approx 11\%$.
                                                                          
\begin{figure}
\includegraphics[height=2.65in]{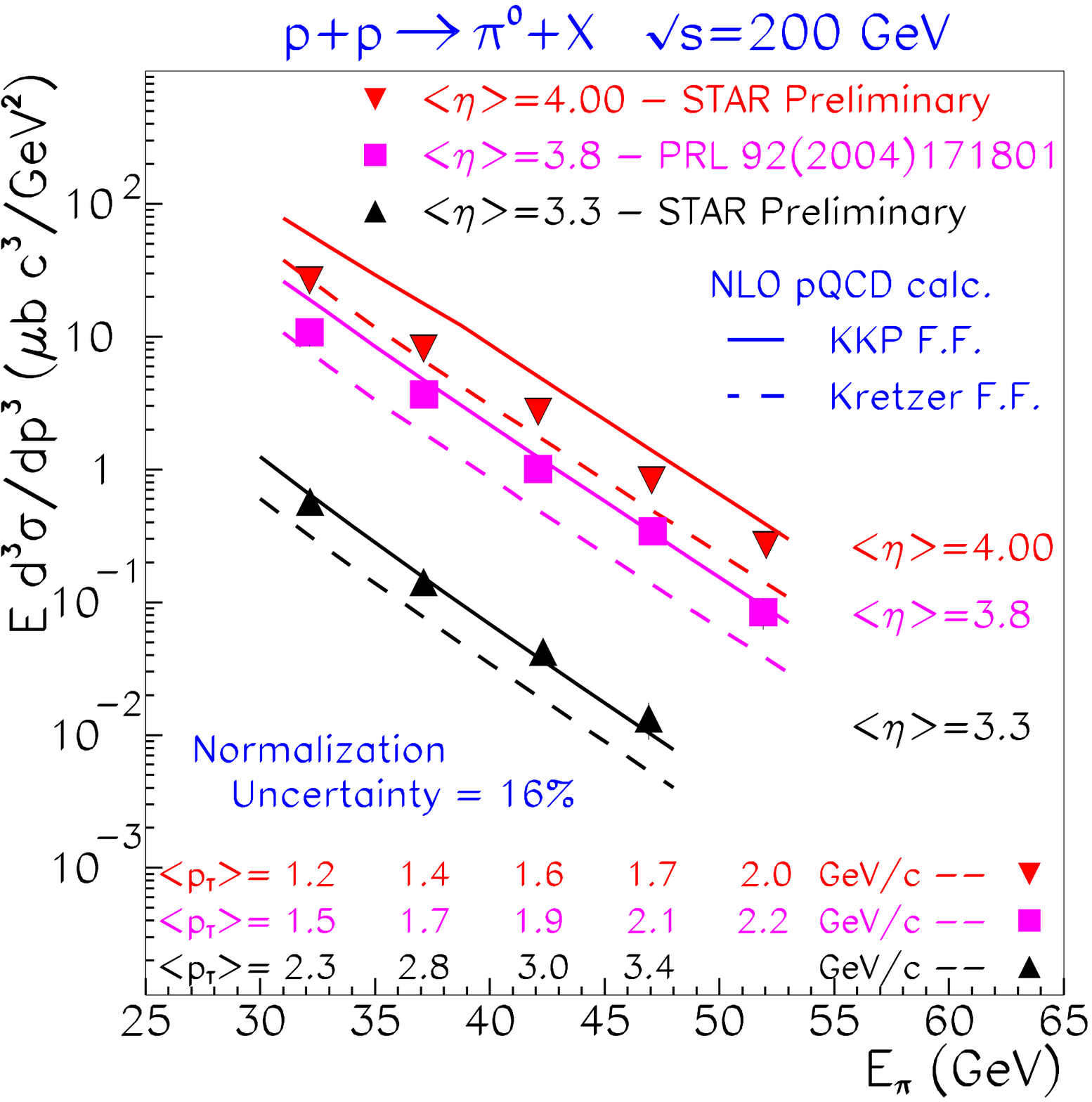}
\includegraphics[height=2.65in]{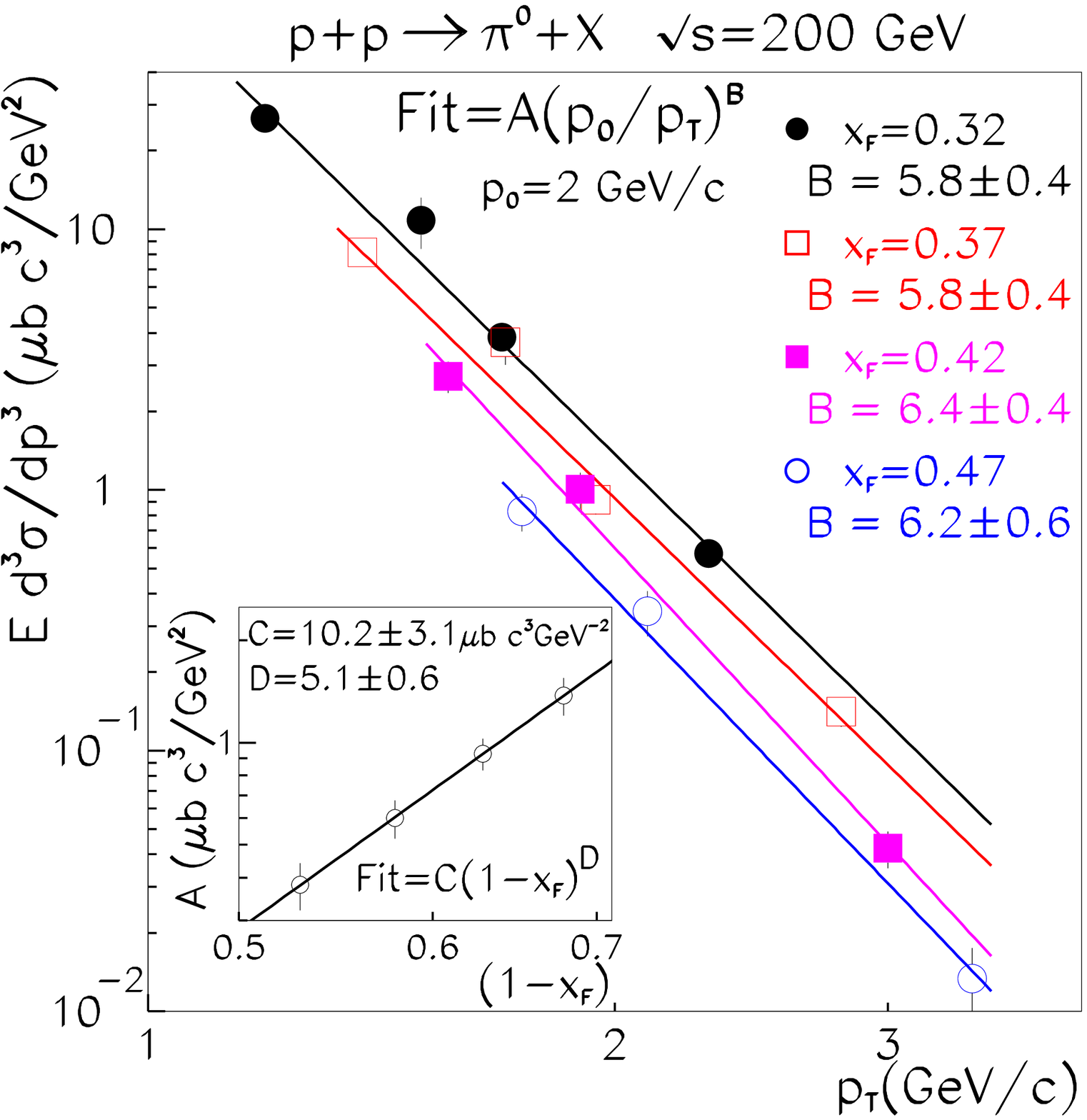}
\caption{
Inclusive $\pi^0$ production cross section for p+p collisions.
The error bars, often smaller than the points, combine the 
statistical and point-to-point systematic errors.
(Left) Versus pion energy ($E_\pi$) at fixed pseudorapidity 
($\eta$).
The curves are NLO pQCD calculations using two sets of 
fragmentation functions (FF).
(Right) Versus transverse momentum ($p_T$) at fixed Feynman-$x$ 
($x_F$).
Curves here and in the inset are fits to the data of the form
shown on the plots.
(Inset) Versus ($1-x_F$) at $p_T=2\,$GeV/c.
%The x symbol is for negative hadrons using $\eta=3.07$ to 
%calculate the value of $x_F$, and is not included in the fit.
\label{fig:inclusive}}
\end{figure}

The cross sections for $p+p\rightarrow\pi^0+X$ at
$\langle\eta\rangle =3.3$, 3.8, and 4.00 are presented in 
Fig.~\ref{fig:inclusive} (left) \cite{STARpi0,moriond05}.
The $E_\pi$-dependent systematic error at 
$\langle\eta\rangle =4.00$ is $10-20\%$, dominated by the 
energy calibration of the FPD.
The absolute $\eta$ uncertainty contributes 11\% to the 
normalization error \cite{moriond05}.
The curves are NLO pQCD calculations using the KKP and Kretzer 
sets of FF's differing primarily in the gluon-to-pion FF, 
$D^\pi_g$.
At $\eta=3.3$ and 3.8, the data are consistent with KKP.
As $p_T$ decreases further, the data begin to undershoot KKP
and approach consistency with Kretzer.
This is consistent with the trend seen at midrapidity
\cite{PHENIXpi0}.
At low $p_T$, the dominant contribution to $\pi^0$ production
becomes gg scattering, making $D^\pi_g$ the dominant FF 
\cite{kretzer}.
In Fig.~\ref{fig:inclusive} (right), the cross section is shown 
versus $p_T$ at fixed Feynman-$x$ 
($x_F\approx 2\,E_\pi/\sqrt{s}$), and versus $(1-x_F)$ at 
$p_T=2\,$GeV/c (inset).
%The BRAHMS cross section for $h^-$ production \cite{BRAHMS} is 
%also plotted, using $\eta=3.07$ to compute $x_F$, and is 
%consistent with the $x_F$ dependence fit from the $\pi^0$ 
%data.
As was reported at the CERN ISR \cite{isr}, the forward rapidity 
yields are rapidly changing functions of both $x_F$ and $p_T$.
%The results demonstrate good control of the systematics, in 
%that the data were accumulated with different electromagnetic 
%calorimeters over multiple years with different readout 
%electronics.

\begin{figure}
\includegraphics[height=2.0in]{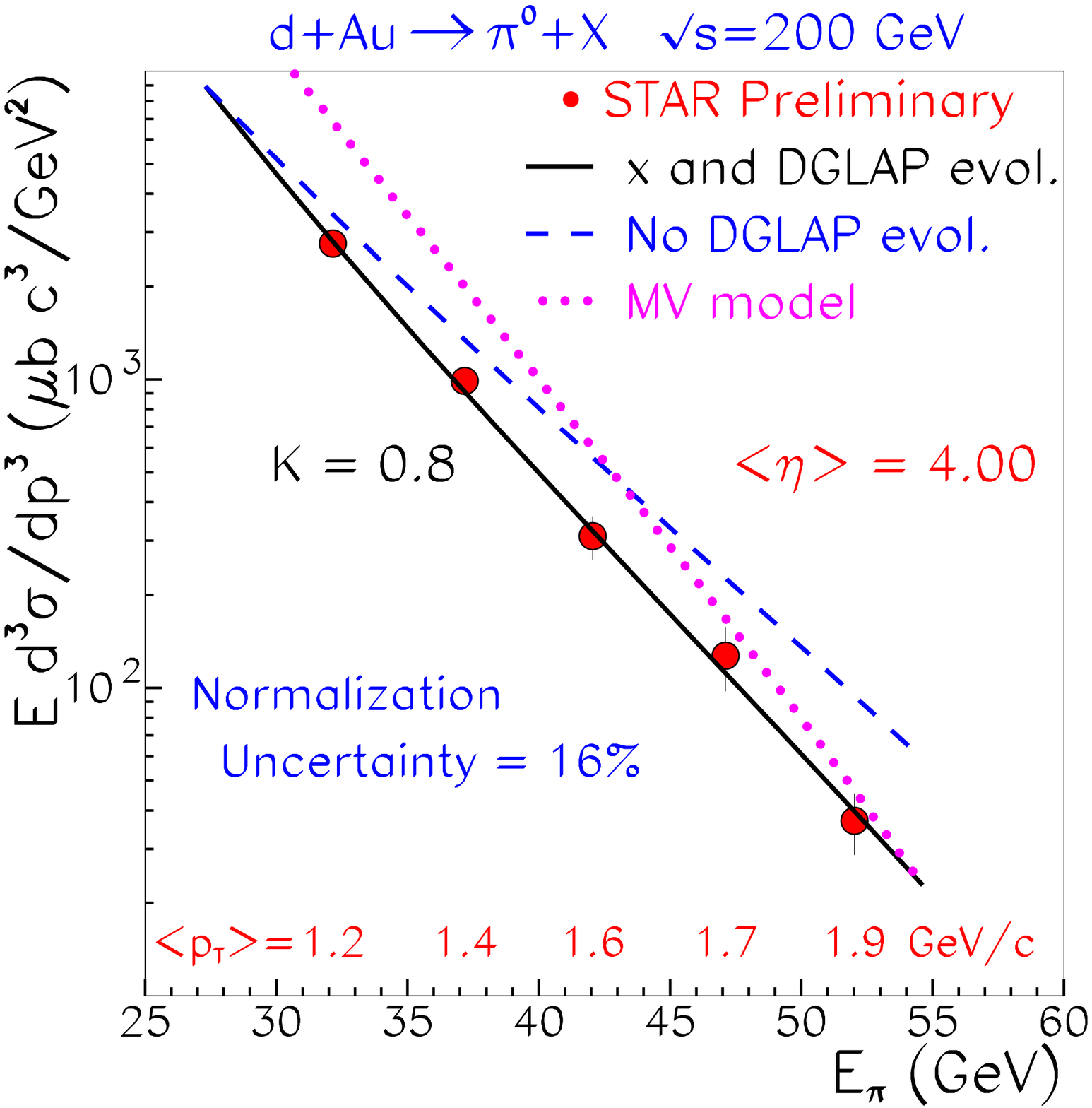}
\includegraphics[height=2.0in]{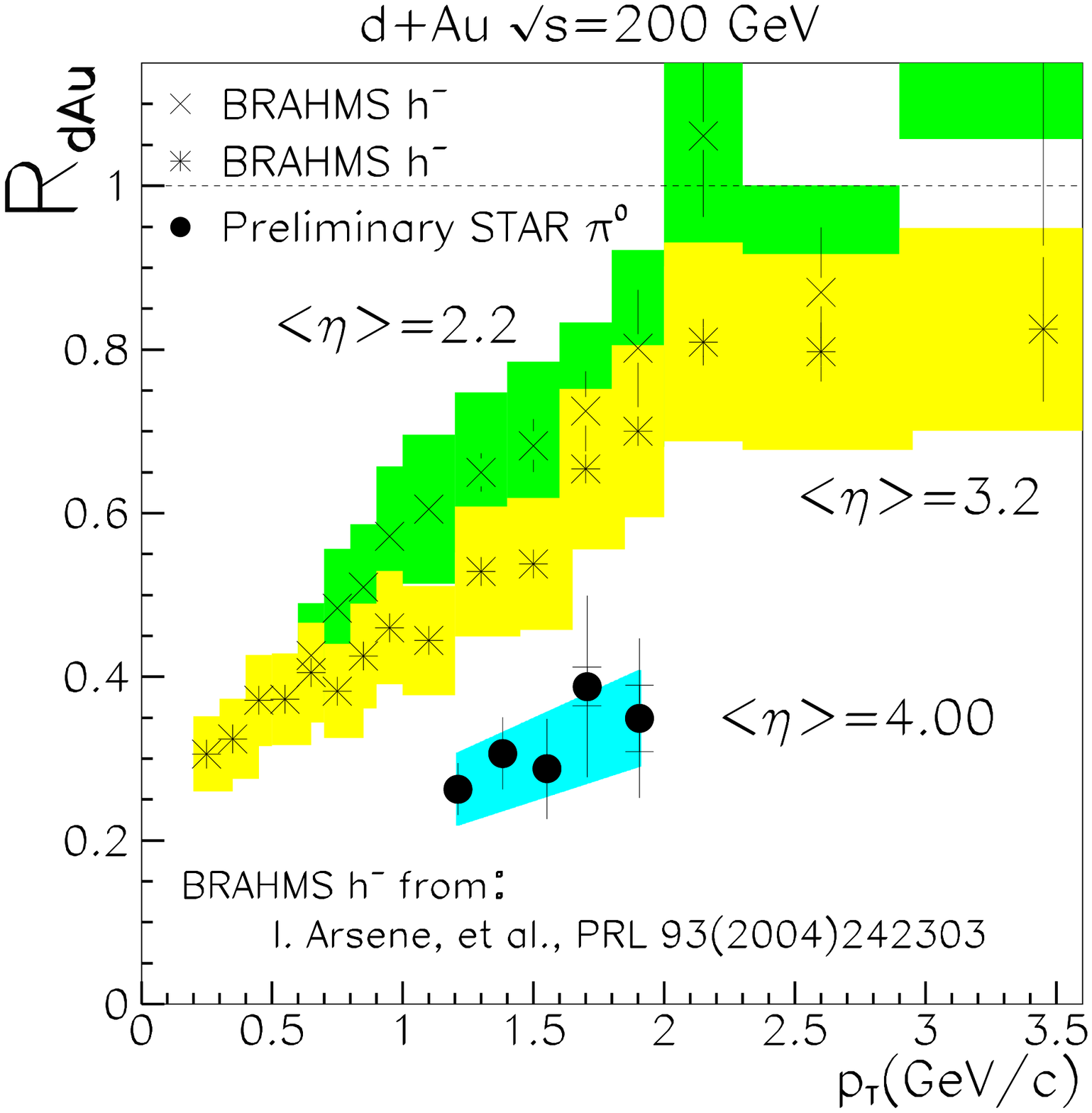}
\caption{
(Left) 
Inclusive $\pi^0$ production cross section for d+Au collisions,
displayed as in Fig.~\ref{fig:inclusive}.
The curves are from models which treat the Au nucleus as a 
CGC, normalized with a $p_T$-independent $K$-factor.  
(Right)
Nuclear modification factor for minimum-bias d+Au collisions 
versus $p_T$.
The solid circles are data for $\pi^0$ mesons.
The inner error bars are statistical, the outer combine these
with the point-to-point systematic errors, and the shaded band is 
the normalization error.
The x's and stars are data for $h^-$ at smaller $\eta$.
These error bars are statistical, while the shaded boxes
are the point-to-point systematic errors.
\label{fig:rda}}
\end{figure}

The cross section for $d+Au\rightarrow \pi^0+X$ at
$\sqrt{s_{NN}}=200\,$GeV and $\langle\eta\rangle =4.00$ is 
presented in Fig.~\ref{fig:rda} (left).
The $E_\pi$-dependent systematic error is $\approx 20\%$, 
dominated by the background correction because,
on average, 0.5 more photons are observed in d+Au than in
p+p collisions per event with $>30\,$GeV detected in the 
FPD.
The curves are LO calculations from 
Ref.~\cite{cgc-cross-section}, using CTEQ5 PDF's and KKP
FF's convoluted with a dipole-nucleus cross section
which models parton scattering from a CGC in the Au 
nucleus.
The ``MV'' and ``No DGLAP'' curves neglect QCD evolution 
of the Au wave function and the PDF/FF, respectively.
The difference in the slopes with and without evolution is 
greater than the slope change from LO to NLO for
$p+p\rightarrow \pi^0 +X$ at $\eta=3.8$.
The calculations in Fig.~\ref{fig:rda} are normalized by a 
$p_T$-independent $K$-factor of 0.8.
This is smaller than the $K$-factor used to normalize the 
theory
to the $d+Au\rightarrow h^-$ yield at the nominal value of 
$\eta=3.2$, in the same direction as the renormalization 
needed to scale the NLO calculations with KKP FF to 
$p+p\rightarrow\pi^0$ data at $\eta=4.00$.
Note from Fig.~\ref{fig:inclusive} that a change of 
$\Delta\eta\approx 0.05$ at these kinematics results in 
$\Delta\sigma/\sigma\approx 35\%$ at fixed $x_F$.
The slope of the $\pi^0$ yield is consistent with the 
calculation where the PDF and FF evolve \`{a} la DGLAP, 
and includes small-$\xbj$ evolution of the Au wave function. 

Nuclear effects on particle production are quantified by the 
nuclear modification factor, $R^Y_{\rm dAu}$, which can be
defined for minimum-bias events by the ratio,
\begin{equation}
R^Y_{\rm dAu} = \frac{ \sigma_{\rm inel}^{pp} }
                 { \langle N_{\rm bin} \rangle 
                   \sigma_{\rm hadr}^{dAu}}
            \frac{ E d^3\sigma/dp^3 (d+Au\rightarrow Y+X)}
                 { E d^3\sigma/dp^3 (p+p\rightarrow Y+X)}.
\end{equation}
We adopt $\sigma_{\rm inel}^{pp}=42\,$mb for the inelastic 
p+p cross section, while the non-elastic d+Au cross section, 
$\sigma_{\rm hadr}^{dAu}=2.21\pm0.09\,$b, and the mean 
number of binary collisions, 
$\langle N_{\rm bin}\rangle=7.5\pm 0.4$, are determined by a 
Glauber model calculation \cite{STARdau}.
Normalization systematic errors mostly cancel in the ratio.
At $\eta\approx 0$, $R^{\,h^\pm}_{\rm dAu}\stackrel{>}{_\sim}1$,  
with a Cronin peak at $p_T>2\,$GeV/c \cite{STARdau}.
In contrast, at 
$\langle\eta\rangle=4.00,\ R^{\,\pi^0}_{\rm dAu}\ll 1$, as 
seen in Fig.~\ref{fig:rda} (right).
The decrease of $R_{\rm dAu}$ with $\eta$ is qualitatively 
consistent with models that suppress the nuclear gluon 
density 
\cite{cgc,coherent,shadowing,recombination}.
$R^{\,\pi^0}_{\rm dAu}$ is significantly 
smaller than a linear extrapolation of $R^{\,h^-}_{\rm dAu}$ 
to $\eta =4$, consistent with expectations of isospin 
suppression of $p+p\rightarrow h^-+X$ \cite{gsv}.
  
%%%%%%%%%%%%%%%%%%%%%%%%%%%%%%%%%%%%%%%%%%%%%%%%%%%%%%%%%%%%%%%%%%%
%\section{Summary}
%%%%%%%%%%%%%%%%%%%%%%%%%%%%%%%%%%%%%%%%%%%%%%%%%%%%%%%%%%%%%%%%%%%
In summary, inclusive yields of forward $\pi^0$ mesons from p+p
collisions at $\sqrt{s}=200\,$GeV are consistent with NLO pQCD
calculations.
The cross sections show rapid variation with both $x_F$ and 
$p_T$.
The d+Au yield is consistent with a model which treats the Au 
nucleus as a CGC for forward particle production.
Comparisons with other models will be interesting to perform.
Normalizing to equal numbers of binary collisions, the d+Au yield
is significantly smaller than the p+p yield.
The $\eta$ dependence of the reduction is consistent with models
which suppress the gluon density in nuclei, in addition to
exhibiting isospin effects at these kinematics.
Additional measurements at different centralities and with other 
final states will help elucidate the cause of the suppression. 
In addition, both di-hadron correlation data and quantitative 
theoretical understanding thereof are needed to facilitate 
tests of a possible CGC.

\end{document}